# USAGE OF LIQUID METALS IN THE POSITRON PRODUCTION SYSTEM OF LINEAR COLLIDER

A.Mikhailichenko, CLASSE, Ithaca, NY 14853

*Abstract:* In this publication we collected descriptions of some installations with liquid metals which could be used for high-energy colliders, ILC particularly, for the purposes of targeting, collimation, cooling, collection of secondary particles etc. Some important components of the system with liquid metals, such as pumps, nozzles, windows, and the fluid dynamics in the Lithium lens are described also.

## OVERVIEW

Usage of liquid metals is a rare event in accelerator physics in general. However as the energy density deposited by beams in various targets increased, it is inevitably forces usage of liquid metal substance. First, liquid metals allow higher temperature rise during the beam pulse, second, liquid substance quickly restores its shape and condition after beam pulse and, third, high Z of metal atoms helps in conversion of one type of beams into another[1]. This is typical for the beam targeting for sure as well as for the ion stripping as well. So the liquid metal is possible to use as a conversion target and as the heat evacuator at the same time.

The positron production system of ILC [1] uses hard undulator radiation (~20*MeV* at first harmonic) caring an average power up to 200 *kW* for irradiation of ~$0.5X_0$–thick target followed by a short-focusing magnetic lens. Although only ~16% of this 200 *kW* power deposited in a beam target[2], this deposition occurs in a small volume, so the beam target design becomes a challenging exercise. Right now the baseline is a spinning wheel made from Titanium alloy. In contrast, we considered here the target which uses a liquid Bismuth-Lead alloy in comparison with the spinning Ti or W target. For focusing of positrons we suggested usage of a compact lens where the liquid Lithium serves as a conductor *and* as a coolant[3]. Liquid Lithium (or Bi/Pb or In/Ga) alloy used as a coolant of Pyrolytic Graphite in a gamma beam absorber. High–energy collimator of the primary beam, where the aperture hole formed by a centrifugal force in a spinning liquid metal is another example considered here.

Cooling of objects with liquid metals has a long history. In the beginning of the 50's USA and USSR begin development of Nuclear Power Facilities for the nuclear submarines. In the USSR the works begin in 1952 with Lead Bismuth eutectic alloy [2]. In USA originally the Sodium was chosen as a coolant and experimental submarine one in series of "Seawolf" has been constructed[4].

---

[1] For example, positron pair production rate by the gamma-beam in a target depends on ~$Z^2$.
[2] This circumstance allows installation of few targets in series and collection of secondary particles from each target for further combining in a longitudinal phase-space [12].
[3] Magnetic flux concentrator is a baseline of ILC positron collection system now.
[4] SSN-575 was the third ship of the United States Navy to be named for Seawolf and the second nuclear submarine [3].

Na/K coolants were in use for nuclear reactors with fast neutrons. However extreme chemical activity makes usage of this coolant problematic in civil installations. One positive property of Na/K alloy is that it remains liquid at room temperature, in contrast to Bi/Pb, see Fig.1.

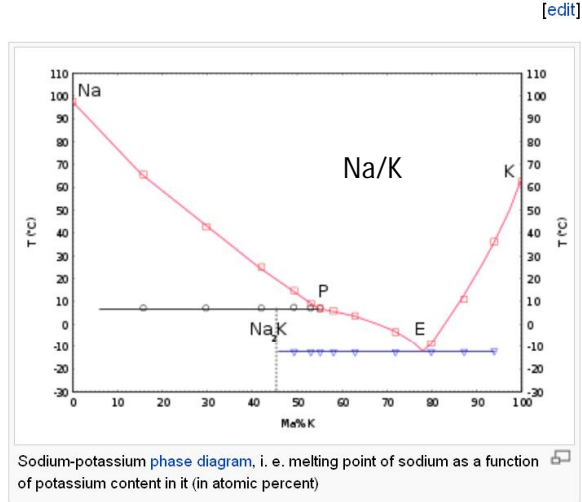

**Figure1.** Fragment of article from Wikipedia.

Usage of Bi/Pb alloy and Mercury demonstrated much more reliability in practice so we concentrated on these metals naturally. Mercury is a toxic agent as a vapor also, but at least it does not demonstrate chemical activity in a properly arranged environment. Other important property of Mercury is that it, as a Na-K alloy, it remains liquid at room temperature. One can find a mentioning of usage of Mercury vapor as an active agent in a turbine [4]. Liquid Mercury jet is also in use in the CERN PS beam as a target [5], [6]. Some other coolant liquids are represented in a Table 1 below (all parameters are functions of temperature, see Table 3 also).

*Table 1. Coolant liquids*

|  | H$_2$O | Lithium | Pb-Bi (55wt%/45wt%) | Hg | Ga |
|---|---|---|---|---|---|
| $r$, g/cm$^3$ | 1.0 | 0.534 | 8.94 | 13.56 | 5.9 |
| $c_p$, J/g/°C | 4.1813 | 3.58 | 0.197 | 0.1395 | 0.37 |
| T$_{melt}$, °C | 0 | 180.54 | 125.9 | -38.83 | 29.76 |
| T$_{boil}$, °C | 100 | 1342 | 1670 | 357 | 2204 |
| $l_{Xo}$, cm | 36.08 | 152.1 | 0.709 | 0.48 | 2.11 |
| Latent heat, kJ/g | 2.26 | 21.2 | 0.86(Pb) | 0.294 | 3.6 |
| Nucl.int.length, cm | 83.3 | 133.6 | 17.6(Pb) | 14.58 | 23.92 |
| Ionization, MeV/cm | 1.992 | 0.875 | 12.7(Pb) | 15.31 | 8.1 |

Finally all heat will be carried out by water, so there is desire to keep the final temperature in a system below 100°C. For the purposes of cooling one can suggest single, dual or even triple cooling-contour systems, Fig.2.

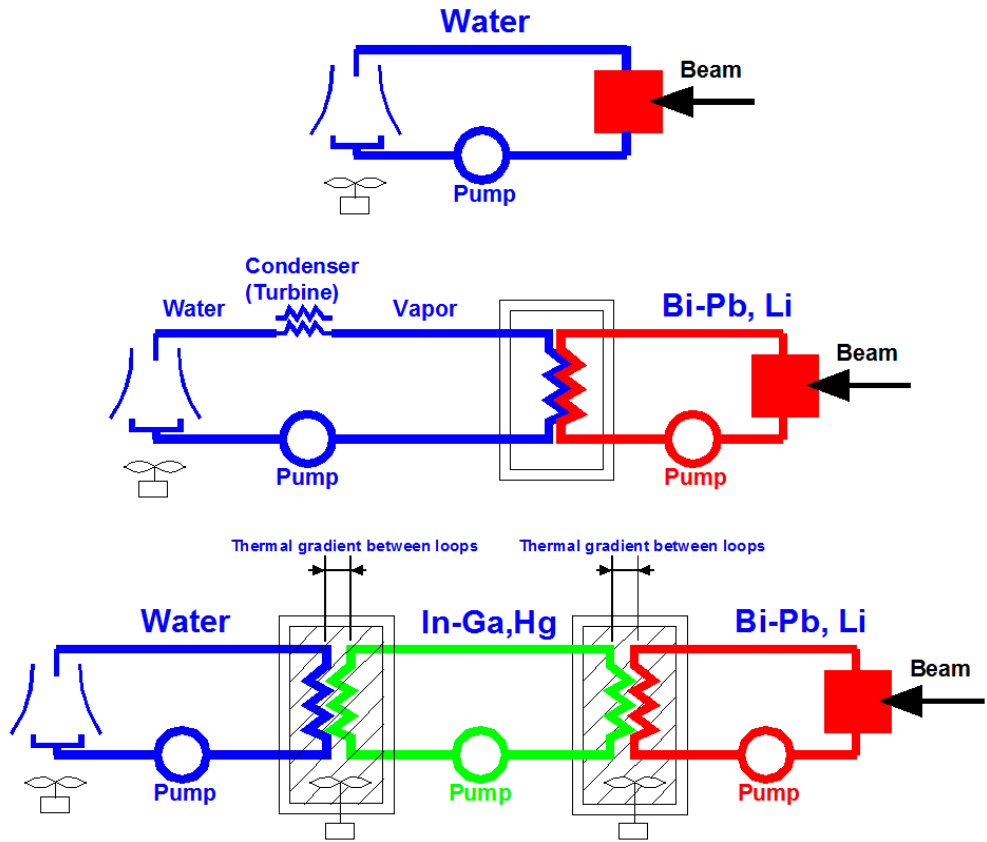

**Figure 2**. Single-contour cooling system, at the top; double-contour cooling system in the middle; triple-contour cooling system at the bottom.

The turbine can be used as a condenser. The condenser can be incorporated into the chilling tower, for beam with high power. In a 2-contour system (middle in Fig.2) the contour with water should be considered as a high pressure one.

Isotopes of Mercury are stable, except artificially created $^{194}$Hg, which decays $^{194}$Hg à $^{194}$Au in ~444 years.

Gallium has better performance when mixed with Indium. Gallium metal price is approaching 1000$/kg, so the system containing 3 liters of Gallium will have a weight ~17.7kg and will cost ~18k$, which is acceptable. Addition of Indium with its ~750$/kg will reduce the price, proportionally to percentage of Indium in the alloy. Savings are not drastic, however.

Lithium metal price ~$64/kg is low at this scale. Mercury metal trades at 700$/34.5 kg; 34.5 kg represents so called flask=76 lb. So from the point of pricing the cost is acceptable for a typical system.

Latent heat of vaporization plays important role, when the rise of temperature of cooling liquid is so high, that this liquid becomes transformed into vapor. Lithium demonstrates the best heat capacity per gram while the temperature rises up to the boiling point. Indeed, the Gallium demonstrates the best heat capacity *per volume*, while its temperature rising up to the boiling point.

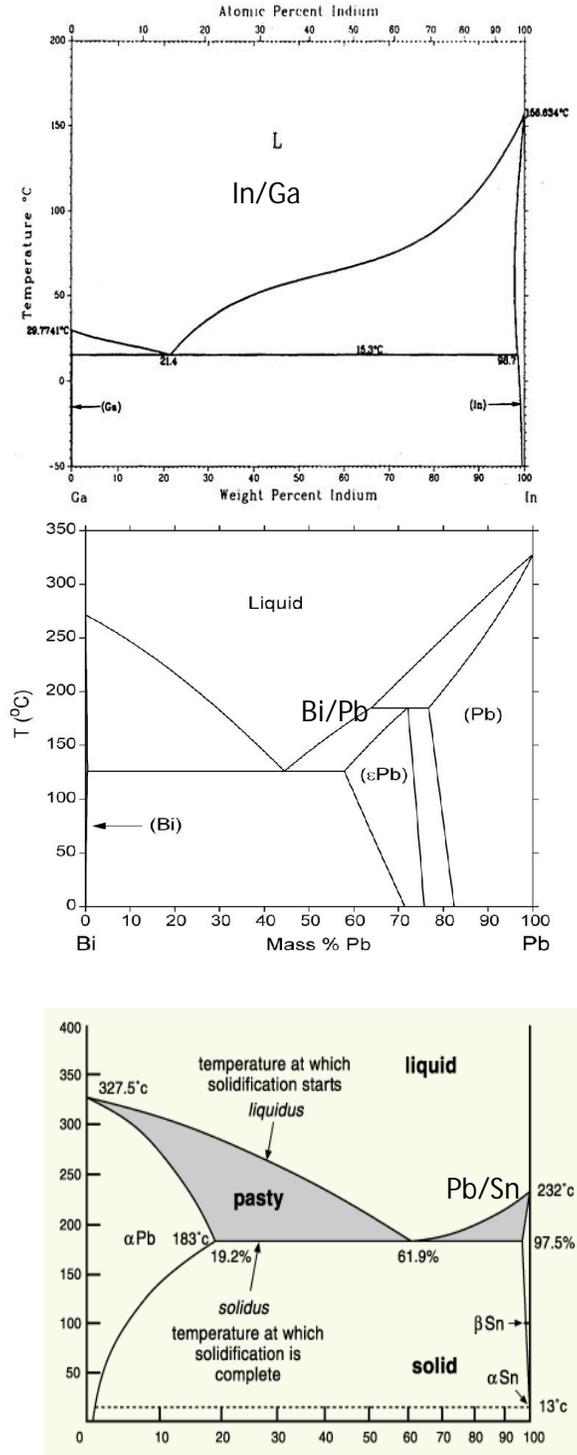

**Figure 3.** Phase diagrams for mostly interesting cooling metal liquids. In/Ga –at the top, Bi/Pb –at the center; Pb/Sn–at the bottom.

Lithium should be kept in a kerosene (Lithium is lighter, than a kerosene, which has specific mass density $\rho \approx 0.82 g/cm^3$) or in a paraffin oil ($\rho \approx 0.868 g/cm^3$). In/Ga alloy does not require any special conditions, Hg –pretty much also.

## PUMPS FOR LIQUID METALS

As the temperature of melted metals considered does not exceed 200°C, a lot of different systems can be suggested for pumping. Some of them are practically the same as for pumping oils or water, but some are using the fact, that metallic substance is a good conductor of electricity.

### Magneto-hydrodynamic pump (MHP).

The concept of this kind of pump is represented in fig 4. Here the metal duct with rectangular cross-section $A \times h$ immersed into magnetic field directed alongside $h$, and the current is running across the metal flow rectangular to the direction of magnetic field, Fig 4.

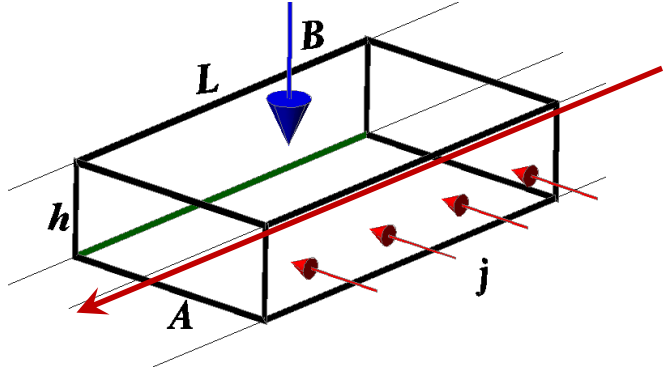

**Figure 4.** To the pressure jump developed by magneto-hydrodynamic pump. Segment of metal duct with length $L$. Metal flows in direction of red arrow (along $L$).

The gradient of the pressure can be found from the following equation
$$\vec{\nabla} P = \vec{j} \times \vec{B} \approx |\vec{j}| \times B, \qquad (1)$$
where $\vec{j}$ is the current density, Here a circumstance was taken into account that the current flow is rectangular to the magnetic field direction. We are interesting in a pressure jump along the segment of Lithium duct having length $L$ (Fig.4), so $\Delta P = |\vec{j}| B \times L$. Substitute here $|\vec{j}| = \dfrac{I}{L \times h}$, where $I$ stands for the total current running across Lithium, one can obtain
$$\Delta P = \dfrac{I \cdot B \cdot L}{L \times h} = \dfrac{I \cdot B}{h}, \qquad (2)$$
which is not a function of length $L$. $L$ should be chosen so that the current density is not big, however. Here the pressure drop $\Delta P$ measured in *Pascal*, $B$ in *Tesla*, $h$ in *meters*. For example, if $I=400\,A$, $h=4mm=0.004m$, $B=1T$, then $\Delta P=10^5$ Pa $\approx 1 atm$.

Sometimes, for simplicity, the magnetic field and current generated by the same AC source (through appropriate transformer), so the direction of magnetic field and direction of current are synchronized automatically even for periodic electric powering.

***The flow meter*** works with the same components as the pump. Here *the voltage* across the Lithium flow is measured. Electric field appeared across the flow can be calculated as

$$\vec{E} = \vec{v} \times \vec{B},\qquad(4)$$

where $\vec{v}$ stands for the velocity of Lithium. So the E.M.F. will be

$$E.M.F. = v \cdot B \cdot A.\qquad(5)$$

For example, if $v=1$ m/s, $B=1$ *T*, $A=10$ *cm*$=0.1m$, then E.M.F.$=0.1V$, i.e. the macroscopic value, which could be measured pretty accurately.

The duct wall captures some fraction of the current, however if made from metal with high resistivity (Titanium, StSteel), the percent of the current captured by the duct walls can be made minimal.

***Spiral type*** of MHD pump is represented in Fig.5 [6]. Here the length *L* compacted by spiraling it and the current runs from central region to the periphery, basically from the inlet tube to the outlet one with help of joint 5 in Fig.5.

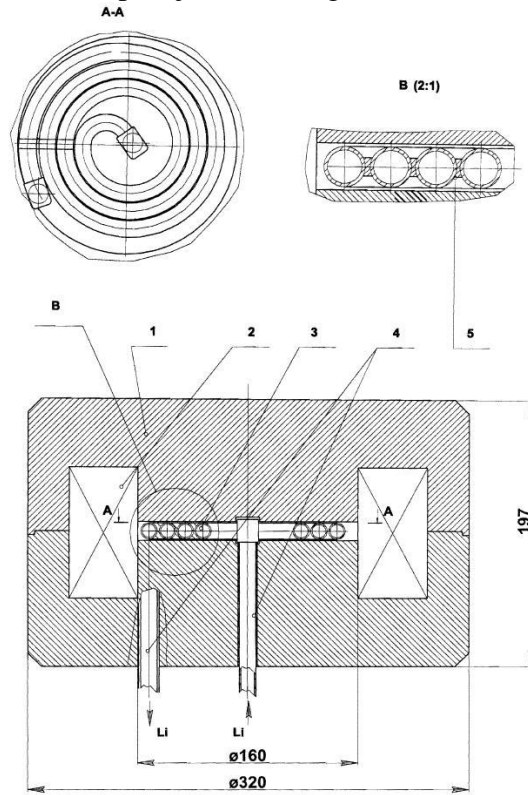

**Figure 5**. Liquid Lithium pump of spiral type [6]. Dimensions are in millimeters. 1 –is the magnet yoke, 2– is the coil 3– is a spiral duct, 4– Li inlet/outlet, 5– are contacting joints.

***Gear pump.***
Our personal preference is a gear pump, however, Fig 6. Lot of high temperature pumps is available commercially. Materials used for gaskets in the high-temperature pumps are Aluflex (-54$^{o}$C-+500$^{o}$C), Blue-Gard (+370$^{o}$C), Fiberfrax (+700$^{o}$C-+1260$^{o}$C), Viton (+220$^{o}$C). High temperature gaskets produced also by Auburn Co. and McNeil Co.

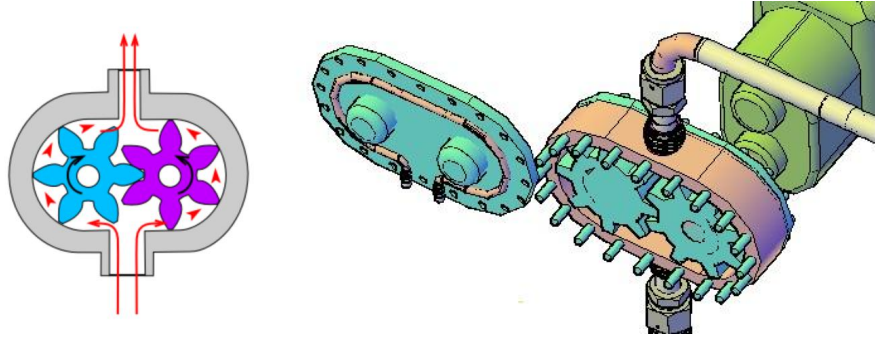

**Figure 6.** Gear Pump.

*Centrifugal pumps* could be used also, but these pumps require higher speed of rotation, as the pressure developed dynamically, so this might be a disadvantage. In Fig.7 a setup is shown which uses a gear or centrifugal pump.

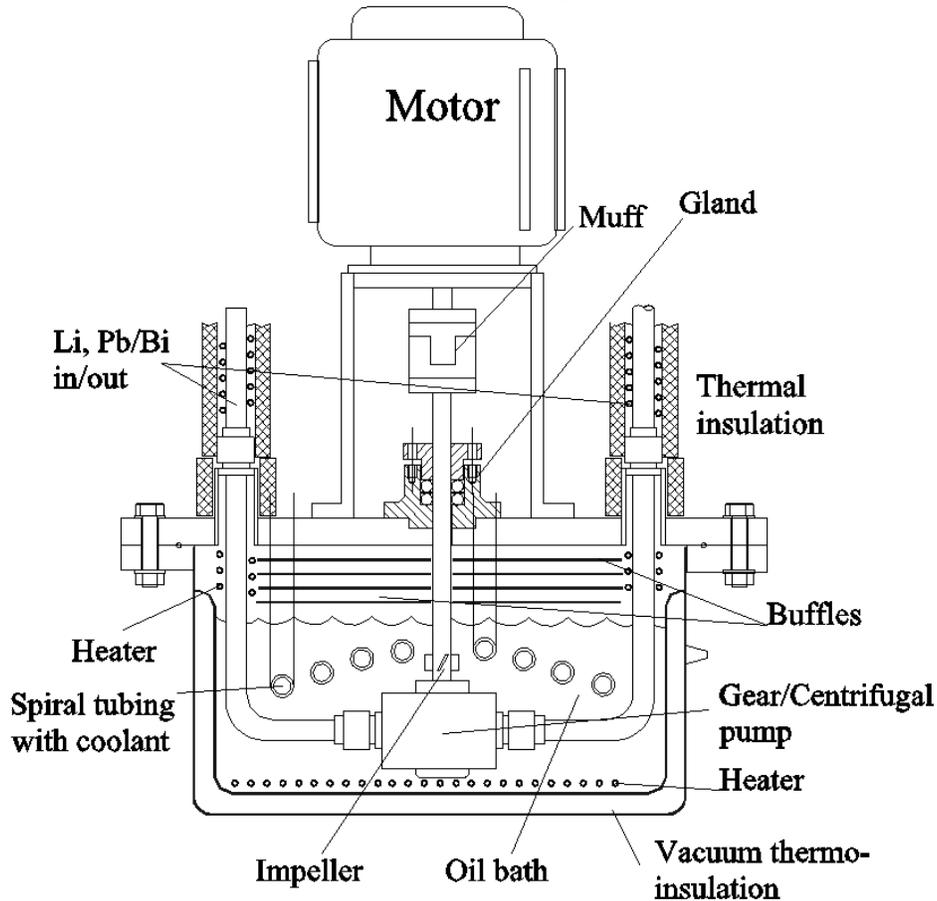

**Figure 7.** Pump setup combined with a heat exchanger. The pump is working for liquid Li, Bi/Pb and for Hg also. In last case the heaters are not required.

In a setup shown in Fig.7, the oil serves as a heart carrier between metal and coolant (water). Temperature of oil could be hold higher (up to 300$^{o}$C), than the melting temperature of Bi/Pb alloy (125°C). Heater serves for melting alloy when the process begins from room temperature. In the case of pumping Hg, the heaters are not required. Oil bath helps in greasing of spinning parts and protects Li from contacts with atmosphere. Impeller serves for stirring of oil. Heat evacuated by coolant running in spiraled tubing. Water with its vapor could be used here as a coolant-so it is a two-contour system.

*Piston type pump.*

This type of pump could be used if periodicity of beam pulsing is not high. This is definitely so for the proton-muon conversion system used for generation of pions and muons further in muon colliders, see for example [5], [6] and for antiproton generation [7], when periodicity is equal to the cyclic sequence of proton beam extracted after acceleration in proton accelerator. One example of such device is represented in Fig.8.

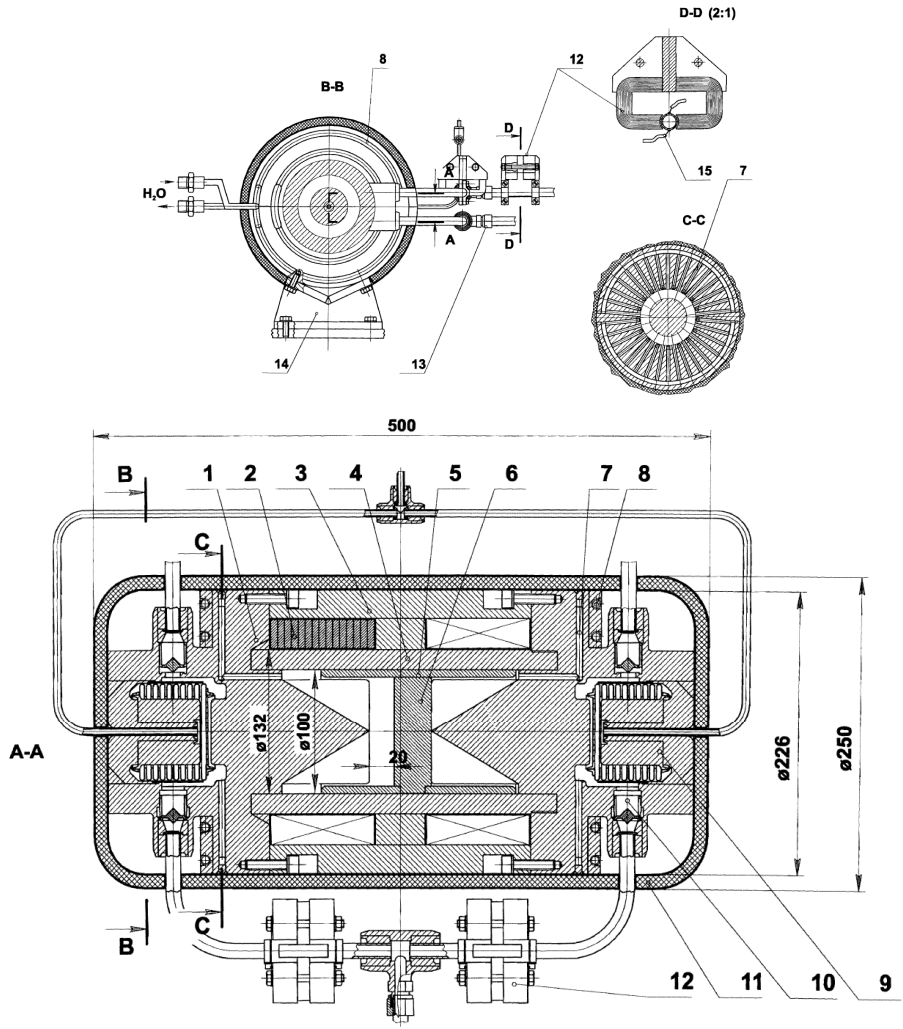

**Figure 8.** Liquid Lithium pump and heat exchanger [7]. 1-face discs, 2-exciting coil, 3-return yoke, 4-inside ferromagnetic wall of cylinder, 5-plunger, 6-ferromagnetic insert of plunger, 7-cooling channels of heat exchanger, 8-cooling water, 9-system to apply and control of liquid Lithium static pressure, 10-switching valves, 11-thermo-insulation, 12-system for measurement of liquid flow rate, 13-joint of liquid Lithium pipes, 14-pump support, 15-contacts for potential measurements.

# BEAM DUMP

*Electron/positron beam dump for intermediate energy.* Such type of dump required in the chain of positron production for absorbing of beam of electrons evacuated from the secondary shower. There electrons and positrons have about the same intensity. The reason for this is the following. The positron created by the photon in a field of nuclei is always accompanied by the electron. When the flux concentrator is used for capturing of secondary particles (positrons), then this device focuses electrons and positrons equally, so they are coming in the first accelerating structure with ~same distribution in transverse and longitudinal phase space, i.e. with maximal energy ~ energy of gamma minus doubled rest energy of electron : $E_\gamma - 2mc^2$ ~ 19 *MeV*. So while the positrons are in the accelerating phase, for electrons this is a decelerating one. So for avoiding deposition of energy in the accelerating structure, after some acceleration in a short structure, the electron and positron beams should be separated. To be able manipulate with low energy electrons theirs energy should not be low, but within few MeV.

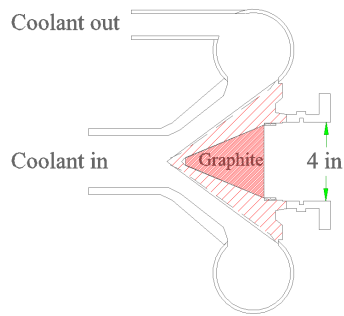

**Figure 9**. The concept of dump for the beam with energy 20-40 *MeV*. Spring pushes graphite cone inside a metallic cone, so the graphite remains in touch with cooled surface. Beam is coming from the right.

Conical shape of absorbent material in Fig.9 -Pyrolytic Graphite (PG)- is useful, as the cone holds the shape even when it becomes hot, just expanding in longitudinal direction. Dimensions of Graphite cone chosen so that it covers all shower, what is of the order of Moliere radius, see [10].

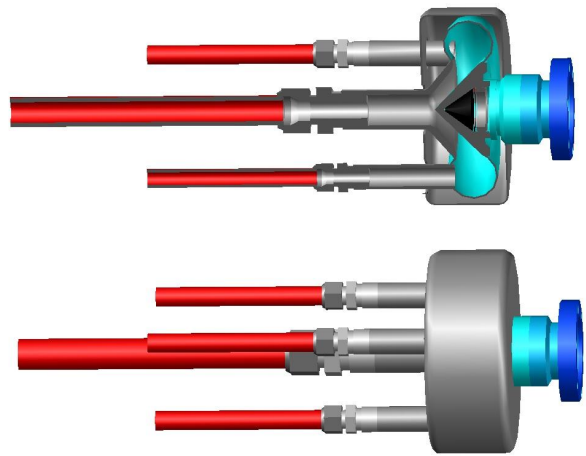

**Figure 10.** 3D view of the beam dump from Fig.9.

Liquid In/Ga or even Li and Hg coolant could be recommended for this system with low energy. Some fraction of the beam, which is not absorbed by graphite, could be absorbed by the coolant itself. For high energy the Bi/Pb coolant cold serve for this purpose. So basically the graphite serves here as a spoiler.

PG is strongly anisotropic material; so orientation of its axes should be chosen so that PG has maximal thermal conductivity in a transverse direction (towards the cold wall.

***Gamma-beam dump and gamma-collimator for ILC*** represented in Fig. 11-Fig.14 below.

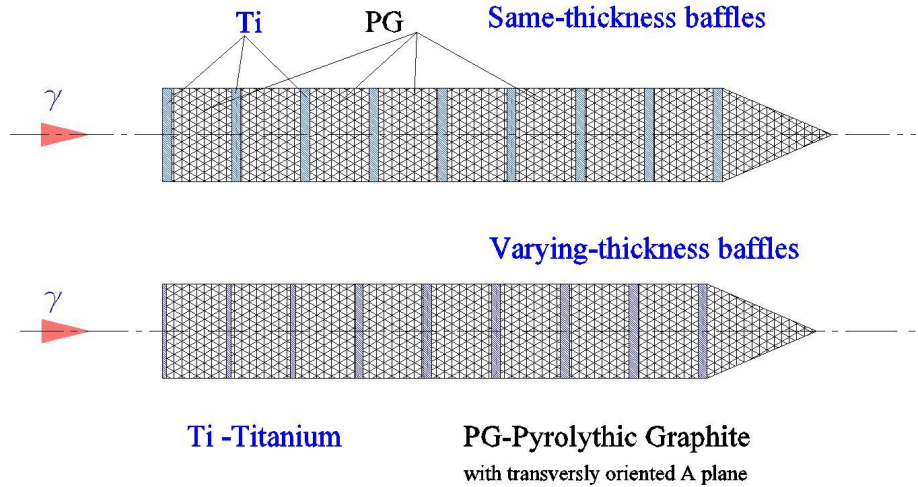

**Figure 11.** Pyrolytic Graphite interlaced by Ti spoilers [10],[17].

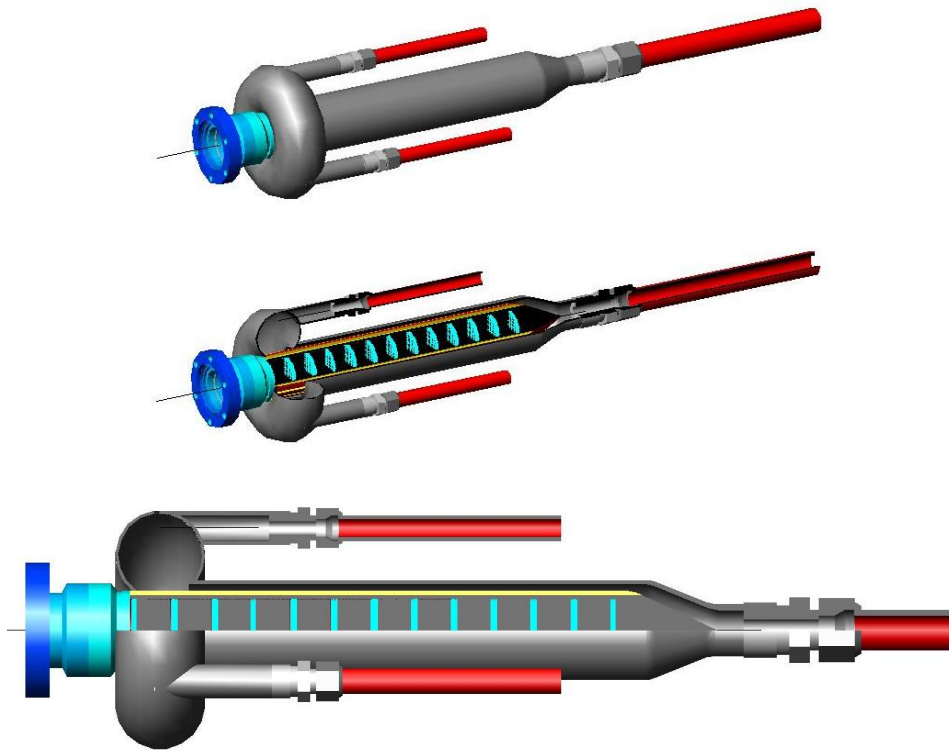

**Figure 12.** Gamma-beam dump; isometric view and cross-section [10].

If the power absorbed is not high, water could serve as a coolant. For higher absorbed average power, In/Ga alloy could be recommended.

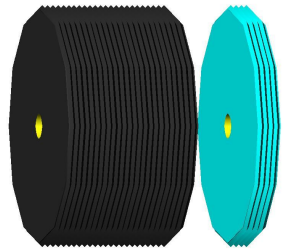

**Figure 13**: PG and Ti discs forming the collimator.

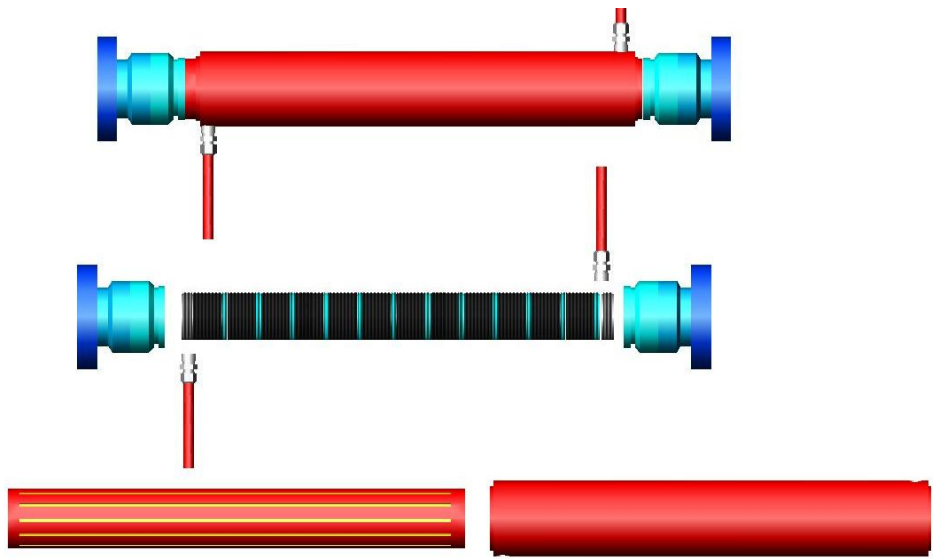

**Figure 14: P**hoton collimator [10].

So basically, the collimator construction is very similar to the construction of gamma-absorber-the only holes identifying the difference. Both Ti spoilers and PG fixed in the tube by thread for better thermal contact allowing some freedom while expanding. Although the collimator normally absorbs the small fraction of beam power, sometimes it should be designed so it should be able to absorb a single shot of full-power beam; the safety chain should turn-off the beam flow after the power deposition becomes critically high (what is definitely so, when the beam hits the collimator directly.

## LITHIUM LENS

Basically a Lithium lens is a Lithium cylinder confined in cylindrical case in such a way, that the feeding current of few tens (or even few hundreds) of kilo-Amps and the beam under focusing are co-propagating *in* the Lithium. Due to low *Z* of Lithium, multiple scattering could much smaller, than the angular spread of secondary particles, so the focusing is a dominating process. For feeding the Lithium lens we can recommend usage of thyristors 5STH 20H4501; each of these allows commutation current up to 80 *kA* at 4.5 *kV*. As the pulse required for ILC is relatively long (~1 *ms*) it is not a problem to form appropriate pulse shape with an artificial line; these line could have three LC contours working at frequencies related as 1:3:5.

Systems with Lithium could be classified as the ones which are using Solid Lithium and the other ones which are using Liquid Lithium. In the last case evacuation of heat in circulating Lithium could be done mostly effectively.

*Lenses with solid Lithium* have very long history, they are developed primarily in BINP, Novosibirsk [14].

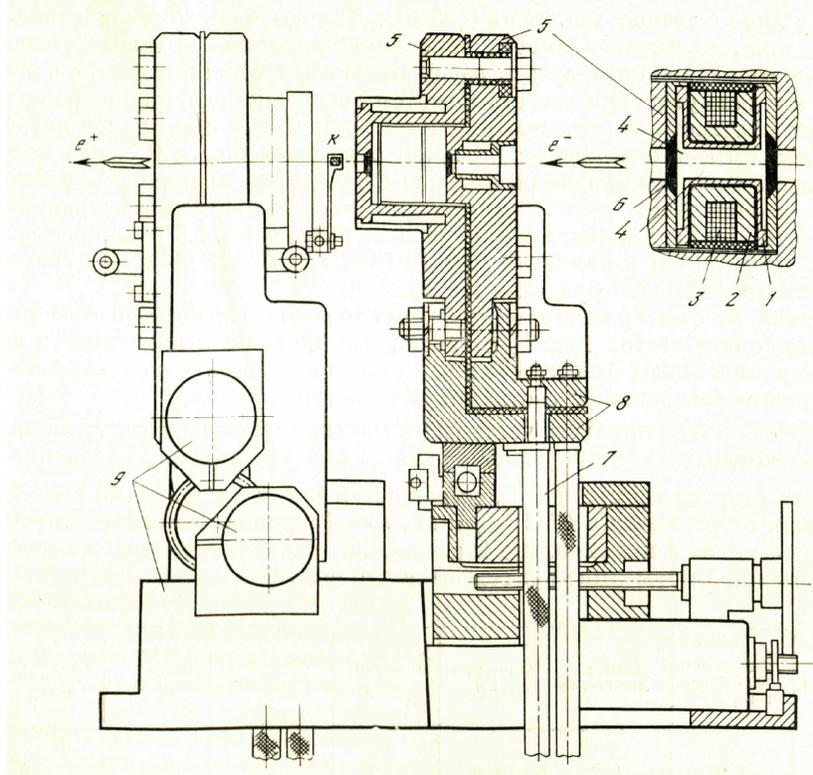

**Figure 15.** Doublet of Lithium lenses with positioning system [14]. 1-Titanium container; 2-Copper clamp; 3-Steel wire bandage; 4-Lithium; 5-current duct; 6-Beryllium windows; 7-flexible current cables; 8-collectors of feeding current; 9-positioning mechanisms. *K*- stands for the W target.

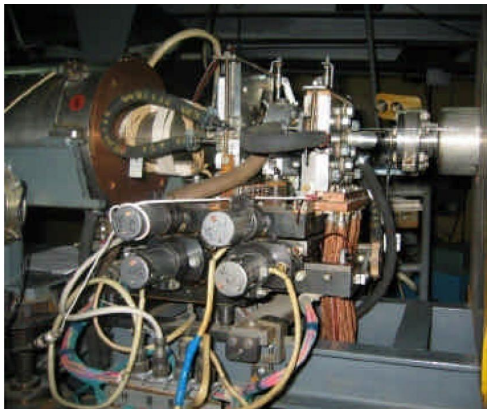 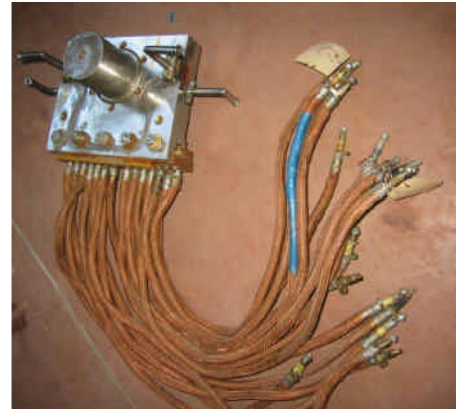

**Figure 16.** Lithium lens setup from Fig.15, left; The lens after ~30 years of operation, right. (Photo courtesy of Yu. Shatunov, BINP)

The lens shown in Fig.15 and Fig.16 were in service at VEPP-2 complex - compact storage ring at BINP.

Another lens used for antiproton production at FERMILAB, also developed in Novosibirsk, is shown in Fig.17. Here the lens is shown installed in a transformer.

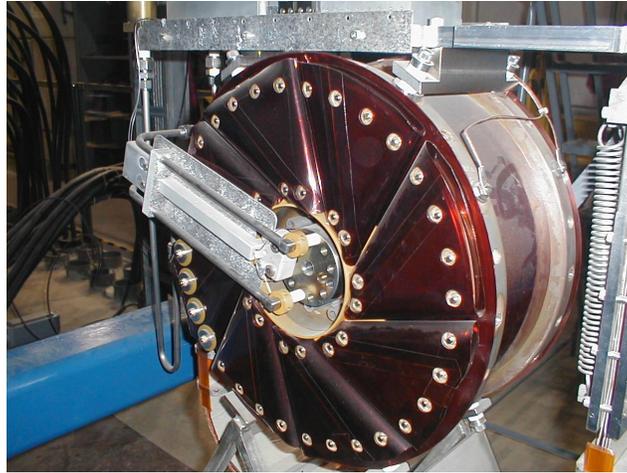

**Figure 17.** Solid Lithium lens at FermiLab. Tubes at the front carry the water coolant.

*Lens with liquid Lithium.* Another type of lens with Lithium is shown in Fig.18.

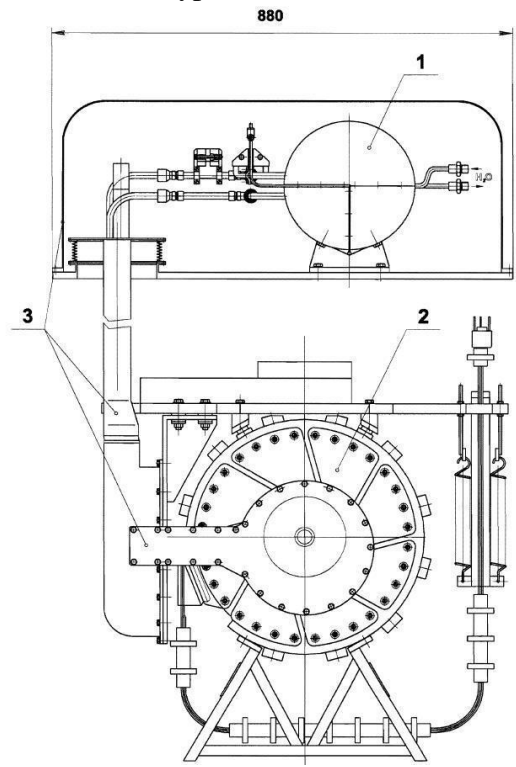

**Figure 18.** Antiproton target station with liquid Lithium [7]. 1-liquid Lithium pump, 2-toroidal transformer, 3-Lithium circuit protection covers. Dimension shown in *mm*.

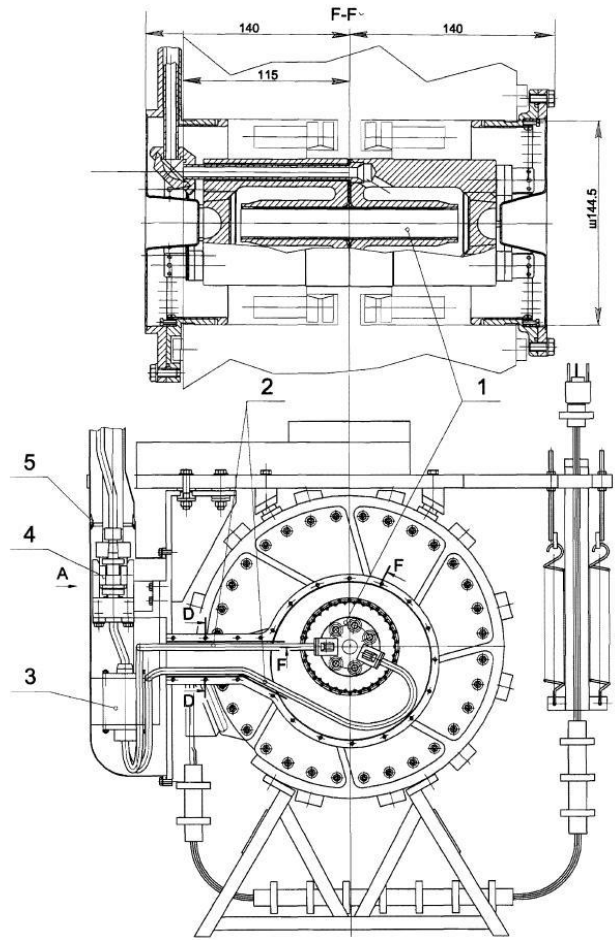

**Figure 19.** Lens with liquid Lithium [7]. It looks similar to the lens with solid Lithium, Fig.17. 1-lens inside the toroidal transformer; 2-connecting tubes; 3-locking valves; 4-system of Lithium circuit remote disconnection; 5-sealed joint Lithium contour protection. Dimensions – in *mm*.

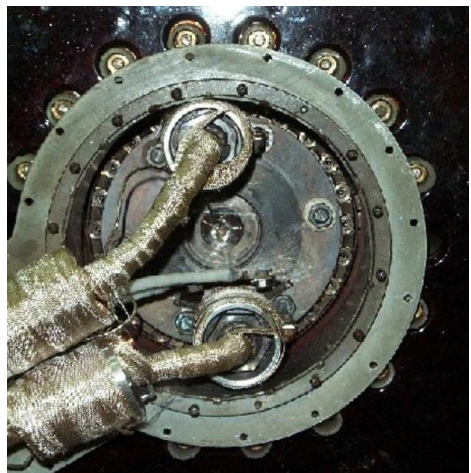

**Figure 20.** Detailed view of Liquid Lithium lens input from Fig.19. [7]

*Liquid Lithium for ILC positron converter.* This comes out as a scaled version of the lens with liquid Lithium developed for proton-antiproton conversion [13].

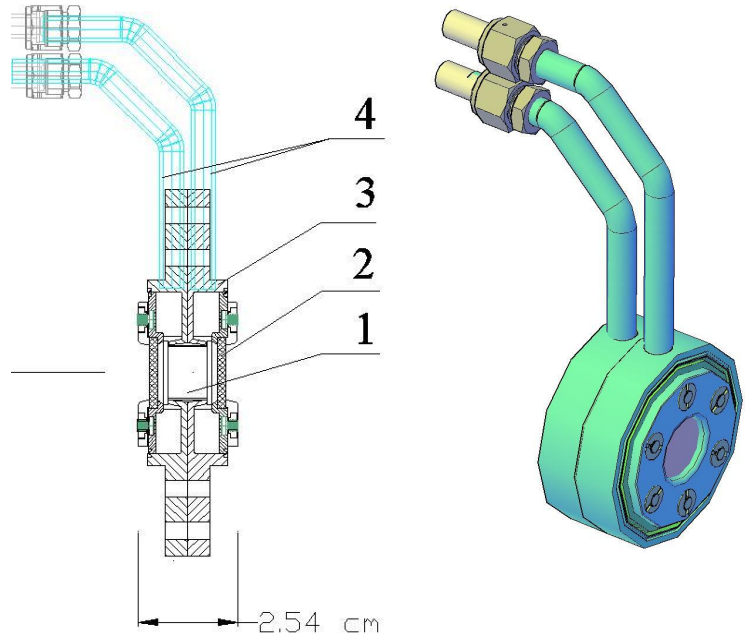

**Figure 21.** Lithium Lens for ILC positron source; extended flanges serve for electrical contact. 1–volume with Lithium, 2–window (Be/BC/BN), 3–electrical contacts with caverns for Li, 4–tubing for Lithium in/out.   At the right –the lens for collet contact attachment (classic).

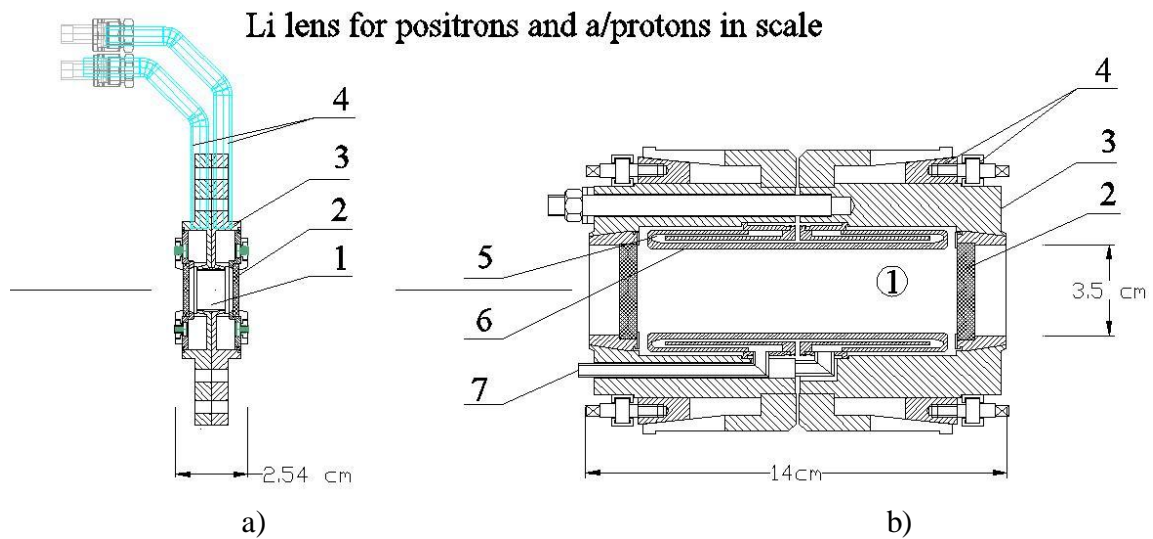

a)                                b)

**Figure 22**. Lithium lens for ILC, a)  and FermiLab antiproton lens b) in comparison [18].  a) 1-volume of LL, 2-Be/BN windows, 3-boady of lens, 4-LL inlet/outlet (left); wedge reinforcement contact (right). 5-is the water cooling channel, 6-Ti wall, 7-is a cooling water duct (one of two).

*Table 2. Parameters of lenses for positrons, antiprotons and for neutrino-factory* [18]

|  | Positrons | Antiprotons | Neutrino factory |
|---|---|---|---|
| Diameter, *cm* | 1.4 | 2-3.6 | 1.8- 6 |
| Length, *cm* | 0.5-1 | 10 | 15 |
| Current, *kA* | <150-75 | ~850 | 500 |
| Pulse duty, *msec* | ~4 | 0.1 | ~1 |
| Repetition rate, *Hz* | 5 | 0.7 | 0.7 |
| Resistance *μΩ* | 32 | 50 | 27 |
| Gradient, *kG/cm* | <65 | 55 | 45 |
| Surface field, *kG* | 43 | 100 | 80-40 |
| Pulsed Power, *kW* | ~720-360 | 36000 | 6750 |
| Average Power, *kW* | ~15-7.5 | 3.6 | 4.7 |
| Temperature gain/pulse, °*K* | 170-85 | 80 | 80 |
| Pressure at axis, *atm* | 75-19 | 400 | 256-64 |

*Windows*

Beryllium is strongly resistant to Liquid Lithium up to temperatures of 500°C. The penetration depth is 0.115*mm/year*. At 300°C its resistance is much higher.

*Table 3. Properties of Li, Be, BN, B4N, W*

|  | Units | Li | Be | BN | $B_4C$ | W |
|---|---|---|---|---|---|---|
| Atomic number, $Z$ | - | 3 | 4 | 5/7 | 5/6 | 74 |
| Yong modulus | $GPa$ | 4.9 | 287 | 350-400 | 450 | 400 |
| Density, $\rho$ | [$g/cm^3$] | 0.533 | 1.846 | 3.487 | 2.52 | 19.254 |
| Specific resistance | $Ohm-cm$ | $1.44 \times 10^{-5}$ | $1.9 \times 10^{-5}$ | $>10^{14}$ | $7.14 \times 10^{-3}$ | $5.5 \times 10^{-6}$ |
| Length of $X_o$, $lX_o$ | $cm$ | 152.1 | 34.739 | 27.026 | 19.88 | 0.35 |
| Boil temperature | °$C$ | 1347 | 2469 | Sublim. at melt | 3500 | 5660 |
| Melt temperature | °$C$ | 180.54 | 1287 | 2973 | 2350 | 3410 |
| Compressibility | $cm^2/kg$ | $8.7 \times 10^{-6}$ | $9.27 \times 10^{-7}$ |  |  | $2.93 \times 10^{-7}$ |
| Grüneisen coeff. | - |  |  |  |  | 2.4 |
| Speed of sound (long) | $m/sec$ | 6000 | 12890 | 16400 | 14920 | 5460 |
| Specific heat | $J/g°K$ | 3.6 | 1.82 | 1.47 | 0.95 | 0.134 |
| Heat conductivity | $W/cm°C$ | 0.848 | 2 | 7.4 | 0.3-0.4 | 1.67 |
| Thermal expansion | $1/°C$ | $4.6 \times 10^{-6}$ | $11 \times 10^{-6}$ | $2.7 \times 10^{-6}$ | $5 \times 10^{-6}$ | $4.3 \times 10^{-6}$ |

---

[1] Total mass of Lithium in ~70*kg* human body is ~7*mg*.

For lenses with liquid Lithium, one should take into account, that the pumping duct shunts the current flow, so adequate measures should be taken. One example could be a non-conducting spinning flow beaker equipped with non-conducting (ceramic) sections of inlet and outlet tube(s).

# LIQUID LITHIUM FLOW IN A LENS.

Dynamic of Lithium flow was considered in [9], [9a] using the model in FlexPDE. In Fig.23 and Fig 24 there are represented the lens duct as it appears in the model.

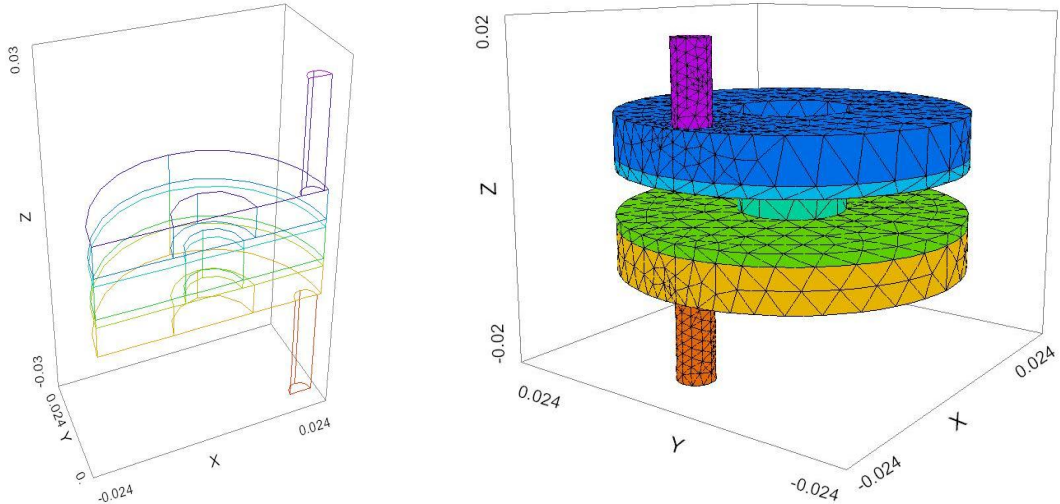

**Figure 23.** Model of Li lens as it appears in FlexPDE.

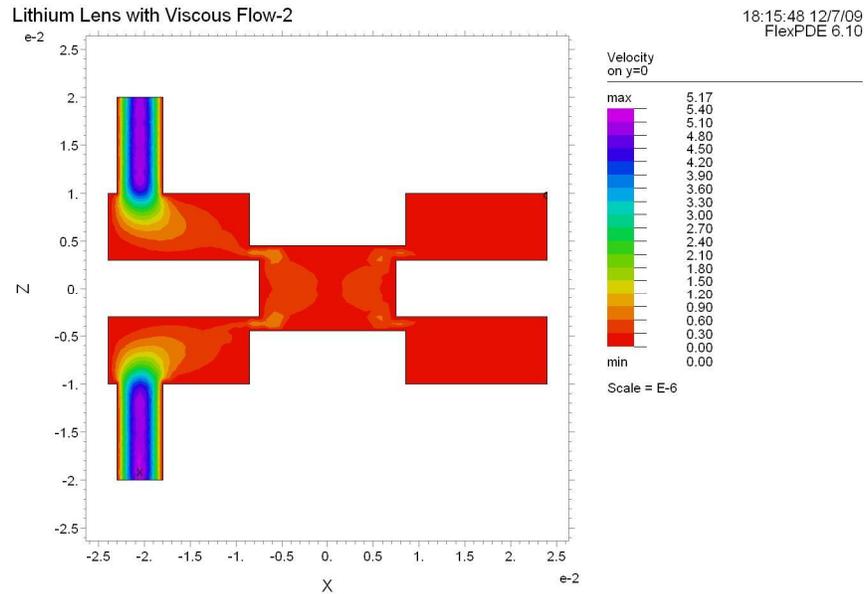

**Figure 24.** Velocity profile without current pulse.

One can see, that asymmetry associated with the inlet/outlet tubes mostly vanished in a region of central Lithium cylinder. Enlargement of toroidal buffer volumes helps in this equalizing.

Vector field of current is represented in Fig. 25. As one could expect, the current density increased in a region of turning around the corner. . Profile of velocity is shown in Fig, 26.

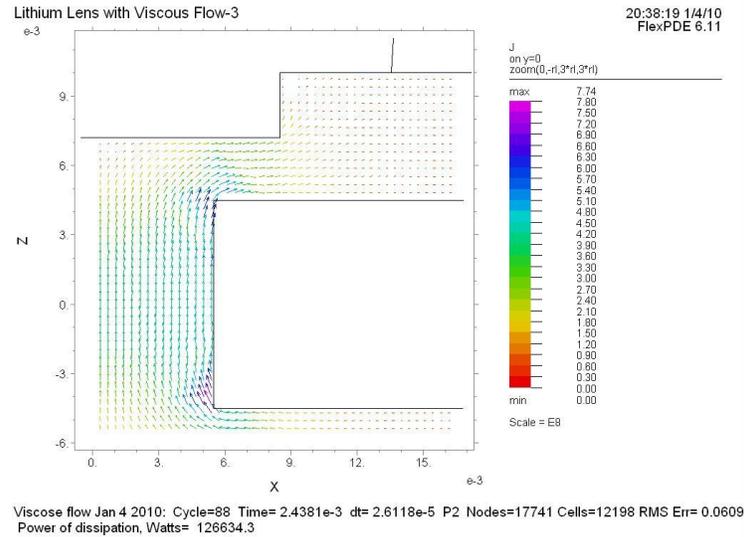

**Figure 25.** Vectors of current.

When the current is running through the body of lens, the flow in a central region demonstrates a motion towards center. This motion forced by magnetic pressure (pinch). In a solid Lithium lens pinch provides detachment of Lithium from the wall. After the pulse ends, the pressure released and the solid Lithium surface hits the wall of confining cylinder what might cause the damage of its wall. In case of liquid Lithium these effects are not manifested, as the empty space immediately filled with new portions of moving Lithium.

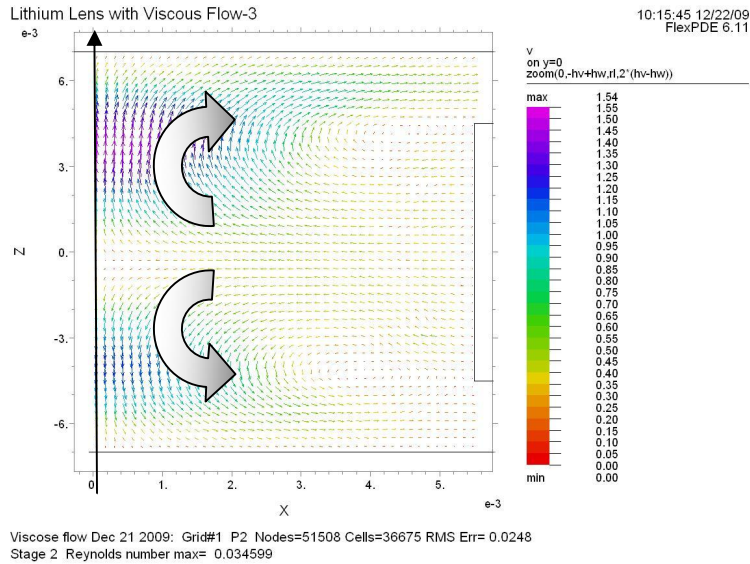

**Figure 26.** Vector field of velocity provided by electrical pulse. Center of lens is marked by the arrow at the left side.

# LIQUID METAL TARGET

Liquid Mercury target was suggested for VLEPP, together with the Tungsten disc which is spinning by this jet [12], [13]. Design, which is represented in Fig. 27, represents a spinning target disc, moved by Hg jet as well as a pure jet target. The Lithium lens is incorporated into the target block.

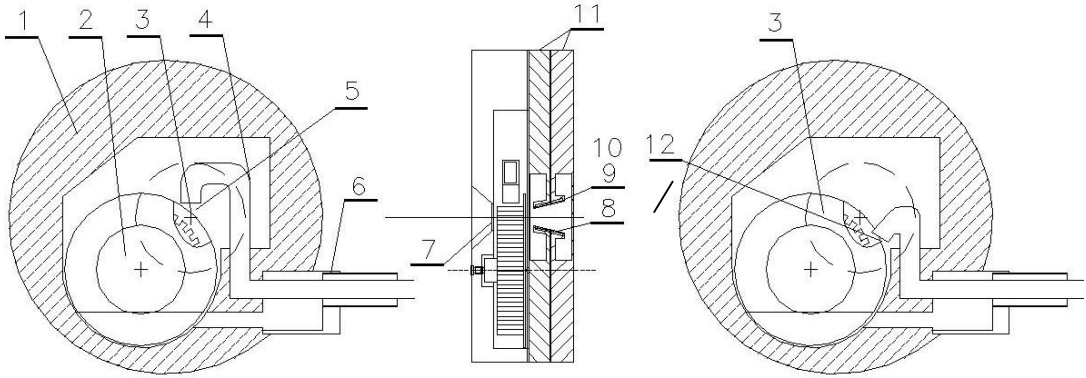

**Figure 27.** Liquid metal target for VLEPP. Variant 1. 1-Titanium case, 2-is the teethed wheel, 3-is the target focusing point, 4-is the nozzle, 5-is the Mercury jet, 6-is the feeding tubes, 7-is secure Titanium foil, 8-is the conically shaped lens, 9-is the volume with liquid Lithium, 10-is Beryllium made flange, 11-are the current leads made from Titanium. Variant 2. 3 is the target focusing point, 12-is the nozzle. Diameter in Lithium cone at the exit of lens is ~1 *cm*.

The liquid metal jet spins the teethed wheel in both variants; this jet carries out the heat also, Thin Titanium disk attached to the target wheel from the side closer to the Lithium lens. This disk serves for protection of window from droplets of Hg in a variant when the jet itself used as a target. Namely in a variant 1 the open mercury jet serves as a target; in the second variant, the jet just spins the Tungsten disk and carries out the heat deposited in a target.

Development of this idea is shown in Fig.28. Here the liquid metal jet spins the teethed metal disk of big diameter and evacuates the heat. The target disc made from W or Ti alloy. This design, as the one from Fig.27 also does not contain any motor for spinning; it also does not require any coolant transfer over spinning axes into the vacuumed volume. Simple feedback system stabilizes the angular speed of rotation.

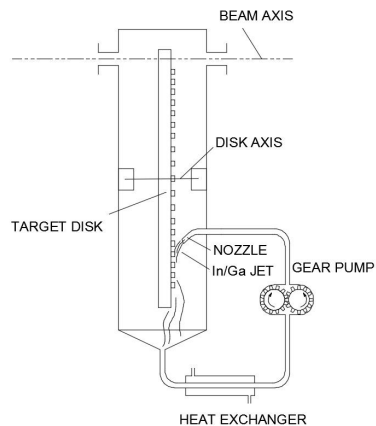

**Fig.28.** Concept of cooling and spinning the target disk of larger scale with metak jet. Diameter of disk~0.5 *m*.

Design of liquid Bismuth/Led target is represented in Fig.29; the jet chamber is shown in Fig.30. The temperature of system should be kept above 200°C, so an adequate thermo-insulation should be provided here.

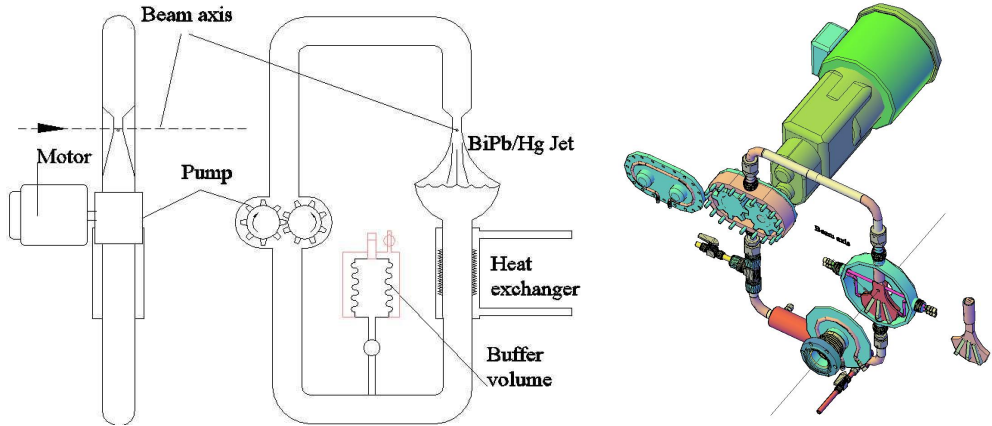

**Figure 29**. Liquid PB/Bi target schematics [11].

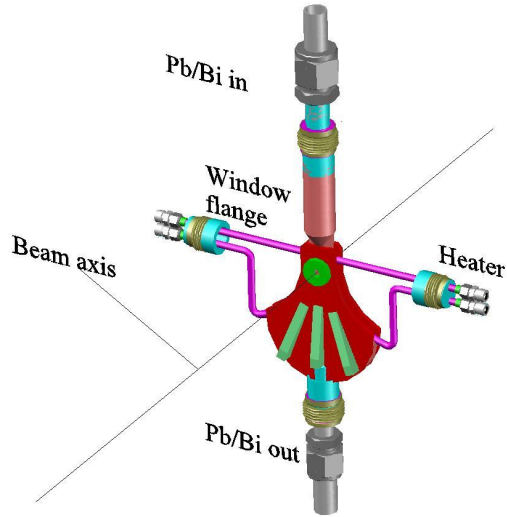

**Figure 30.** The jet chamber. It is shown removed from the vacuumed case.

The jet chamber could be made from Ti (Melt @1668°C ) or Niobium (melt @ 2464°C). Windows cooled by the metal jet itself. Material for windows: $^4$Be; $^{22}$Ti; Boron Nitride- BN ($^5$B$^7$N, sublimates @2700°C). Boron Carbide (B4C), melts @2350 (see Table 3). Jet cross section is (width x thickness)= 1*cm*x0.24*cm*. Jet velocity~10*m/s* provides for 1 *ms* the distance ~1 *cm*. Cross section of jet chamber is represented in Fig.31 a) and b). What is important here is that the Bi/Pb jet after interaction with beam enters expanded cross section, so vaporized component of jet and droplets meet the liquid surface. Windowless jet requires protective shield (as it was suggested in VLEPP targer station, Fig.27) or distance, see Fig.31c. The windowless system for liquid Lithium was considered also in [8]. One should expect that the maximal deflection angle of tiny droplets might be calculated from ratio of jet speed and the speed of sound in a liquid.

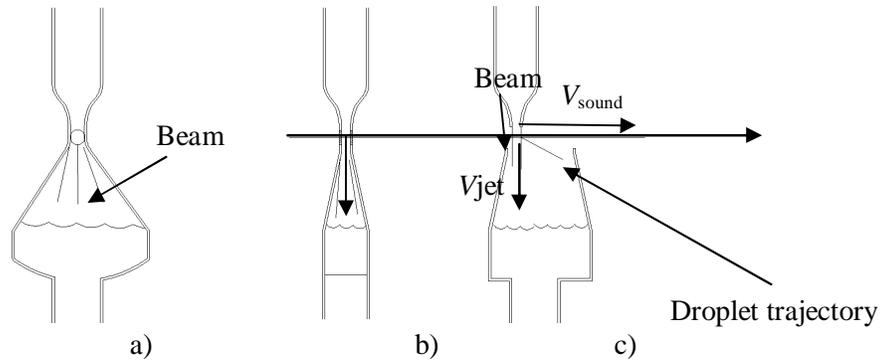

**Figure 31.** a)-front view, b)-side view , c)-side view of windowless system

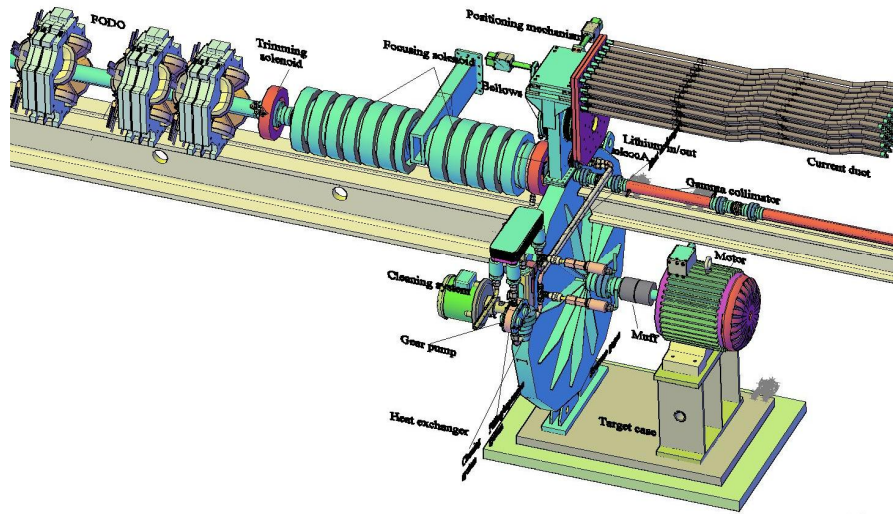

**Figure 32.** The ILC conversion system with rotating target and the liquid Lithium lens. Gamma-beam approaches from the right.

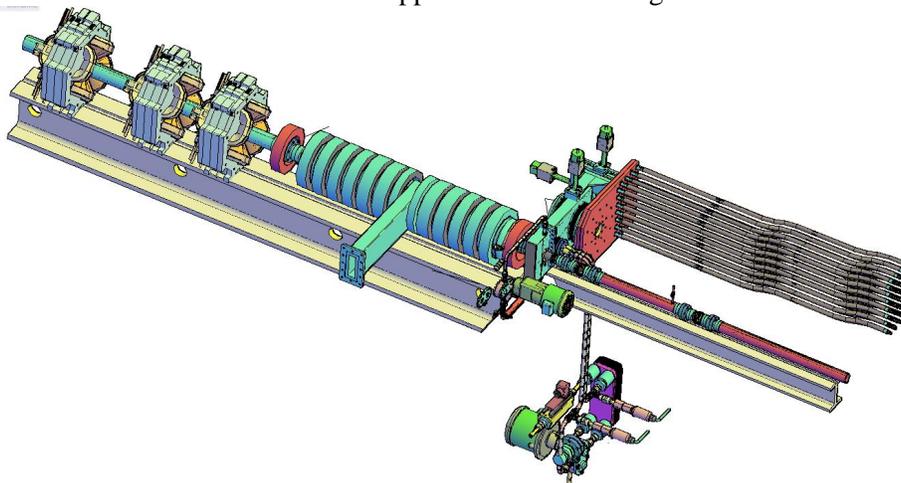

**Figure 33.** ILC conversion system with liquid metal target (Bi/Pb) and the liquid Lithium lens. Gamma-beam approaches from the right.

In contrast with the flux-concentrator as a focusing device (which focuses electrons and positrons equally), the Lithium lens focuses positrons and defocuses electrons (or vice versa), created by target in ~equal quantities. So this peculiarity might be useful for separation of electrons and positrons at the very beginning, [12]. At the entrance of RF section additional collimator should be installed.

## COMBINED TARGET-COLLECTION SYSTEM FOR THE POSITRON PRODUCTION

Lithium lens design described earlier, allows combing the target *and* focusing in the same unit. For this purpose, the incoming flange should be made from Tungsten.

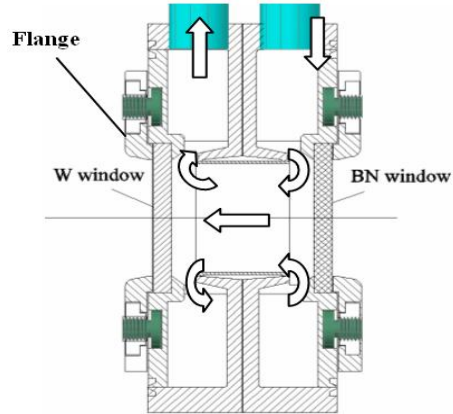

**Figure 34.** (Gamma) beam is coming from the left side.

*Table4. Parameters of conversion system with combined target/lens.*

| | | | |
|---|---|---|---|
| Beam energy, *GeV* | 100 | 150 | 250 |
| Length of undulator, *m* | 220 | 170 | 170 |
| *K* factor | 0.66 | 0.36 | 0.28 |
| Period of undulator, *cm* | 1 | 1 | 1 |
| Distance to the target, *m* | 200 | 350 | 600 |
| Thick. of target/$X_o$ | 0.55 | 0.57 | 0.6 |
| Radius of lens, *cm* | 0.6 | 0.6 | 0.6 |
| Gradient, *kG/cm* | 60 | 60 | 65 |
| Length of the lens, *cm* | 0.7 | 0.7 | 0.7 |
| Current, *kA* | 108 | 108 | 117 |
| Radius of collimator, *cm* | 0.2 | 0.5 | 0.15 |
| Rad, of irises in RF, *cm* | 3 | 3 | 3 |
| Rad of coll. before RF, *cm* | 2 | 2 | 2 |
| Acceptance, *MeV×cm* | 9 | 9 | 9 |
| Energy filter E>, *MeV* | 51 | 54 | 63 |
| Energy filter E<, *MeV* | 110 | 110 | 180 |

| ΔT per train $10^{13}$ $e^-$, $^\circ C$ | 172 | 139 | 270 |
|---|---|---|---|
| ΔT in lens from beam, $^\circ C$ | 18 | 35 | 80 |
| ΔT in lens from current, $^\circ C$ | 90 | 90 | 100 |
|  |  |  |  |
| **Efficiency,** *e+/e-* | 1.52 | 1.57 | 1.52 |
| **Polarization**, % | 54 | 57 | 64 |

Resulting polarization for this system is less than for the one where the target and lens are separated longitudinally (where it reaches ~70% for the same current in a lens). The simplicity of this system might be attractive however.

## HIGH POWER COLLIMATOR FOR THE MAIN BEAM OF ILC

Behavior of electromagnetic cascade at high energy requires substantial length of absorbing material: in longitudinal direction and in a transverse as well, where the size of absorber should be bigger, than the Moliere radius [10]. For collimators, looks, like the only fraction of full energy of the beam might be deposited there. This is true for normal operation. However, the collimator should be able to protect the downstream equipment even if malfunctioning of upstream magnets directs the beam into the wall. This is definitely so for the collimator installed in front of undulator. Undulator has a small aperture, ~5-8 mm and direct hit by train of bunches with ILC population might destroy the undulator permanently. Utilization of collimator considered above, does not solve the problem, as the graphite might be damaged also. So the spinning liquid In/Ga alloy used to form a collimator; that was considered for VLEPP project, where the population of a single bunch was planned to be ~$10^{12}$ in [12], [13].

As the main problem with collimator is destruction of its wall under hit by the train or by a single bunch, the inner surface of collimator should be an open surface of liquid metal. So this liquid metal surface could be formed as result of *centrifugal force*. After direct hit of beam, the metal surface restored pretty quickly, as the droplets of metal meet the opposing surface of the same liquid metal, see Fig.34. Here a Beryllium cylindrical bottle, having holes at both sides, put in spinning around its axis with angular speed ~3000 rev/min. Rotation of this Be bottle arranged with the help of motor through muff and gear, Fig.34. Diameters of the holes in the bottle are chosen close to the aperture of the downstream device, undulator in "out" case. The In/Ga or Hg liquid pumped inside the bottle through the muff, sealed more of less hermetically and freely coming out through the holes in the bottle neck and bottom. Further on, the liquid goes to the heat exchanger with the help of pump, described earlier. As the radiation length of Ga alloy is ~1.5 *cm* and the Moliere radius about the same value the dimensions of such collimator could be very compact.

The thickness of Be walls in regions near input/output holes, should be made thin, so the energy deposited by the beam will be not high; 0.5-1 *mm* thick wall at these locations will be adequate. The heat deposited by the beam in the face and back walls, carried out by liquid metal. The diameter of holes in case of undulator should be slightly smaller, that the aperture of undulator, but the radius of bottle, again, defined by Moliere radius (plus some extra).

As the pressure of vapors of In/Ga alloy is low, there will be not a problem to pump it. (Vacuum pressure of Hg at room temperature~$1.85 \times 10^{-3}$ *Torr*, so differential pumping required)

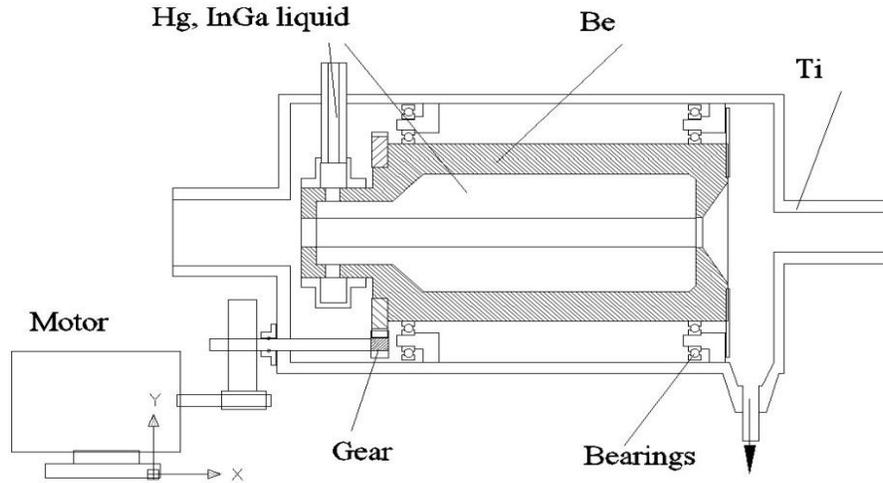

**Figure 35.** Collimator for the beam with high average power. Beam is coming from the right. Flanges are 2 1/8" (or 1.33"), shown is reduced length.

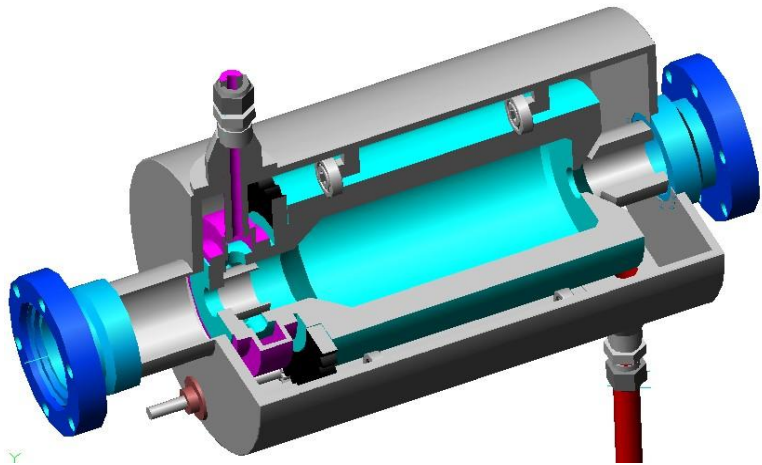

**Figure 35a.** Collimator for the beam with high average power and able to withstand a direct hit by full train of ILC (~2500) bunches. This is an isometric view of Fig.35.

## SUMMARY

We tried to collect some materials about usage of liquid metals in installations for the particle accelerators; however a lot of publications are not illuminated, although they could be found in the referenced publications.

System with liquid metals represents advanced technology for design of targets, collimators and collection optics. Liquid metals have wide area for application in ILC positron/electron source and in the proton targets for muon colliders intensively developing now.

Utilization of Lithium lens allows Tungsten survival under condition required by ILC with $N_e$~$2 \times 10^{10}$ with moderate $K$~0.3-0.4 . Thin high-Z target allows increased creation of positrons and better functioning of collection optics (less depth of focusing is). Liquid metal jet (Pb/Bi or Hg) could be made with variable thickness.

Lithium lens emerges as a well-developed technique, so the lens with parameters required for ILC is guaranteed.

Liquid metal target allows compact design and stability under dynamical load; it could be better, than the one with spinning rim target. Problem of solid target is in dynamic destruction of exit surface due to inertial forces generated by propagating beam; target [5].

Liquid metal jet could be used in a target with spinning W or Ti disk for evacuation of heat and for spinning the target rim (or disk) at the same time.

Highly efficient positron collection system with liquid Lithium lens allows relaxed parameters of the undulator and target.

All calculation made with start to end Monte-Carlo simulation code for conversion – KONN- confirmed that low $K$ factor is possible with focusing by Li lens; $K$<0.4 –which allows making the aperture of undulator ~8 *mm*.

As the field at the entrance (target side) and at the exit side is strictly limited by the surface of the lens, it is possible to use the Lithium lens in a close proximity of spinning target.

Combined Lithium lens- a lens with W entrance window- can work for ILC; for CLIC all parameters are relaxed substantially.

E-166 experiment performed at SLAC [15] confirmed polarization ~80-85% achievable with undulator-based positron production scheme.

We would like to underline finally, that the conversion system for ILC could be realized with solid Ti of W target and with flux concentrator as a focusing device, but usage of liquid metal target and the lens with liquid Lithium allows system to be compact, elegant and less expensive.

---

[5] One can recall SLC target photo with damaged out-surface. As the target performed a complex motion basically resembled a circular rotation (arranged by a kind of single-arm pantograph with circular motion of source point), destruction manifested as a circular groove.